# Efficient Learning Strategy for Predicting Glass Forming Ability in Imbalanced Datasets of Bulk Metallic Glasses

Xuhe Gong [a,b], Jiazi Bi [a], Xiaobin Liu [a], Ran Li[a,*], Ruijuan Xiao [b,*], Tao Zhang [a], Hong Li [b]

[a] *School of Materials Science and Engineering, Key Laboratory of Aerospace Materials and Performance (Ministry of Education), Beihang University, Beijing 100191, China;*

[b] *Institute of Physics, Chinese Academy of Sciences, Beijing 100190, China*

*E-mail: liran@buaa.edu.cn (R. Li), rjxiao@aphy.iphy.ac.cn (R. Xiao).

## Abstract

The prediction of glass forming ability (GFA) and various properties in bulk metallic glasses (BMGs) pose a challenge due to the unique disordered atomic structure in this type of materials. Machine learning shows the potential ability to find a way out. However, the training set from the experimental data of BMGs faces the issue of data imbalance, including the distribution of data related to elements, the range of performance data, and the distribution of sparse and dense data area in each specific system. In this work, the origin of the data imbalance and its impact on the GFA prediction ability of machine learning models are analyzed. We propose the solutions by training the model using the pruned dataset to mitigate the imbalance and by performing an active experimental iterative learning to compensate for the information loss resulting from data reduction. The strategy is proved in Zr-Al-Cu system, and the automated workflow has been established. It effectively avoids the prediction results from trapping into the intensive training data area or from inducing by the data distribution of similar element systems. This approach will expedite the development of new BMGs compositions especially for unexplored systems.

## 1. Introduction

In the era of big data, machine learning has gained tremendous popularity and adoption across various fields, including but not limited to medicine [1], wireless communication [2] and materials science [3-6]. Researchers often use machine learning to analyze and summarize the patterns in large-scale data due to its exceptional processing speed and data mining capabilities. In the realm of materials science, machine learning has been extensively employed by researchers to facilitate property prediction, composition design, and mechanism investigation [7-10]. Especially for materials with demanding experimental preparation processes, high costs and intricate influencing factors, machine learning plays a crucial role in expediting research progress, facilitating data analysis, and more [11,12].

Due to the unique disordered structure, metallic glasses (MGs) have various excellent properties such as high strength and hardness [13], excellent soft magnetic properties [14] and distinct catalytic activity [15], compared with crystalline materials. However, commercial application of MGs is limited by the low GFA. The development of the MGs composition with sufficient GFA and desired properties becomes the key in this filed. The GFA can be directly characterized by the critical cooling rate ($R_c$) or the critical casting diameter ($D_{max}$), and the latter is more frequently taken as the quantity to indicate the GFA because it is easier to measure than $R_c$. Researchers have spent a lot of time searching for BMGs with $D_{max}$ larger than 1 mm, and many high GFA systems have been developed, such as Pd-Cu-Ni-P [16], Mg-Zn-Ca [17], Ti-Zr-Ni-Be [18], etc. Nevertheless, this is still a tiny fraction of the large number of potential systems, and there are still a large number of unknown systems worth further investigation. At present, the composition exploration of most BMGs primarily depends on "trial and error" method. The target composition is verified one by one by the method of rapid melt quenching, significantly diminishing the efficiency of composition discovery. High-throughput experiments utilizing techniques such as magnetron sputtering have been employed to improve the efficiency of experimental verification [19-21]. However, it is important to note that the information obtained from thin film samples produced by magnetron sputtering cannot be directly extended to bulk samples, because the cooling rate in magnetron sputtering is on the order of $10^{12}$ K/s, while traditional BMGs prepared through methods like copper mold casting have cooling rates in the range of $10^3$ to $10^8$ K/s [22]. Despite the possibility of obtaining amorphous thin film samples through sputtering under the same composition conditions, crystallization may still occur during the process of preparing bulk samples due to limitations in cooling rate. In the pursuit of high GFA in BMGs, it becomes imperative to establish efficient composition search methods capable of reducing economic costs and time costs simultaneously. One of the traditional approaches employed by researchers relies on empirical principles such as the eutectic point criterion and Inoue's three principles, or some empirical parameters composed of thermodynamic parameters such as $T_{rg}$($T_g/T_m$, $T_g/T_l$) [23], $\gamma$($T_x/(T_l+T_g)$) [24], $\gamma_m$($(2T_x-T_l)/T_g$) [25] (where $T_g$ denotes glass transition temperature, $T_x$ denotes crystallization temperature, $T_l$ denotes liquidus temperature). Although these parameters have been validated in certain systems, their applicability is limited when developing new materials. On one hand, in order to obtain characteristic temperature parameters such as $T_g$, it is necessary to conduct Differential Thermal Analysis (DTA) or Differential Scanning Calorimetry (DSC) experiments, which requires the preparation of amorphous samples thus these descriptors cannot be obtained during the model prediction stage. On the other hand, the universality of such empirical parameters is also difficult to confirm [26]. It is of great significance to explore a more reliable method for predicting the composition-related properties of MGs, including but not confined to GFA, so as to facilitate quantitative composition design for BMGs.

On this basis, the machine learning method has been adopted to study the different properties of MGs. Daegun You et al. [27] developed an artificial neural networks model that aimed to classify the amorphous/crystalline phases and correlate the excess electrical resistivity of the alloys to the full width at half maximum (FWHM) value of X-ray diffraction (XRD) patterns. Fang Ren et al. [28] successfully distinguished the composition regions capable of forming amorphous films in Co-Ti-Zr and Co-Fe-Zr systems through the iterative combination of high-throughput experiments and machine learning. Jeon et al. [29] established a random forest model to design Ni-based MGs with ideal thermodynamic properties $T_x$ and $T_g$. All of the machine learning models mentioned in above works are trained based on experimental datasets. Because of the inherent disorder in atomic structures of MGs, no high-throughput computational data for amorphous materials is available, making it difficult to utilize simulation data for model training. Consequently, issues related to data distribution and sample imbalance within the experimental datasets become common challenges affecting the quality of model predictions.

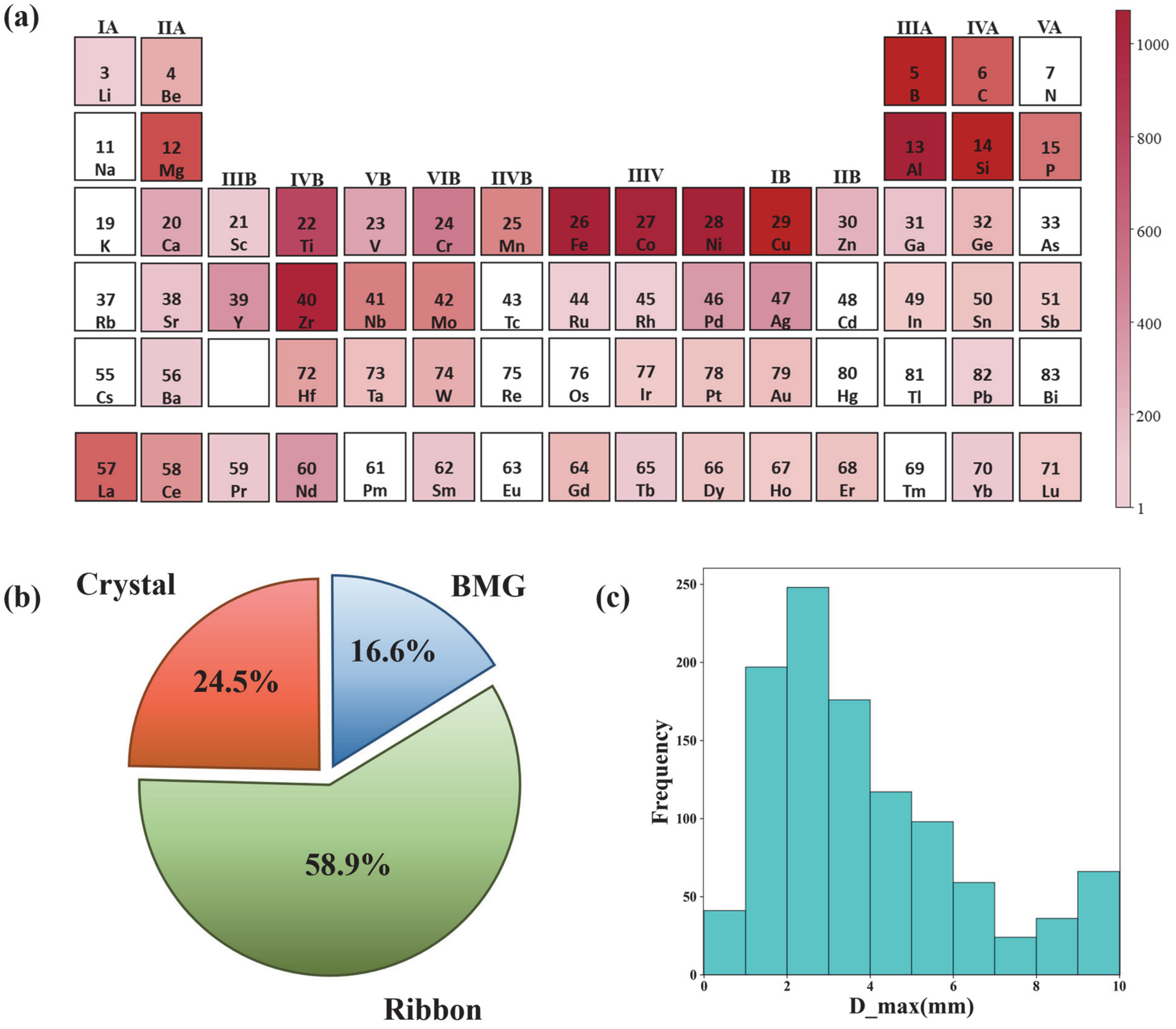

Fig. 1 Distribution analysis of GFA datasets (a) Statistics of element occurrence frequencies in the datasets. (b) Pie chart of label value types. The “Crystal” label represents data that can only be synthesized as crystals, “Ribbon” represents data corresponding to ribbon samples obtained by melt

spin quenching, and “BMG” represents data that can be synthesized as BMGs. (c) Statistical chart of $D_{max}$ of BMGs in the datasets.

The data imbalance manifests in two aspects, one is the imbalanced distribution of data related to elements, the other is the imbalance in the distribution range of performance data since extremely exceptional performance is generally found in a minority of systems. In order to provide a more specific illustration of the data imbalance, we conduct an analysis of the data distribution within a database of MGs by combining datasets from published articles [30-32] and experimental data in our research group [13,33,34]. Taking the values of $D_{max}$ for MGs as an example, firstly, as shown in Fig. 1(a), most of the common alloying elements on the periodic table are covered, among which there are more data of Mg, Al, Fe, Co, Ni, Cu and Zr elements. Secondly, as shown in Fig. 1(b), more than half of the overall database consists of ribbon-like samples, while crystal data accounts for less than a quarter and bulk amorphous data is the least, less than one-sixth. There is a shortage of data on BMGs due to the unclear mechanisms of GFA as well as the lack of efficient techniques for the design and preparation of BMGs. Besides, many experiments that fail to form amorphous structures are not reported, resulting in a relatively smaller amount of crystal data included. In Fig. 1(c), the statistical graph of $D_{max}$ for BMGs data in the database indicates that data with $D_{max}$ larger than 5 mm is generally less abundant compared to data with $D_{max}$ less than 5 mm.

The data imbalance also occurs in each specific system, such as a system with only sparse experiment data points in the entire chemical space but dense amorphous data points near a certain composition. Models trained on imbalanced data in such systems tend to bias prediction results towards the data-rich amorphous composition region. However, in areas with no data, the lack of information doesn’t necessarily indicate the inability to form BMGs. Moreover, it is well known that systems with similar elemental composition often show proximate high GFA regions, which means that the data imbalance within the similar systems affects the prediction of target systems as well. Therefore, the currently available experimental datasets unavoidably suffer from imbalance issues from the elemental distribution and label value distribution within the entire dataset, as well as the non-uniform data distribution in each specific system. When making predictions based on the imbalanced dataset, the results tend to be trapped in some area and the generalization ability is reduced, causing a missing opportunity to discover new composition with high GFA.

Optimization attempts have been made in several research studies to address the issue of imbalanced data of MGs. On the one hand, data enhancement is a good solution for imbalanced label distribution [35]. By adjusting the data sampling method, a uniform distribution of label values can be achieved. On the other hand, some scholars introduce a large number of high-throughput experimental data to dilute the imbalanced feature distribution caused by traditional experimental data [36,37]. However, current high-throughput synthesis of metallic glasses mainly focuses on film samples by magnetron sputtering, which cannot provide the quantitative measure of GFA needed in the regression problem. The impact of training datasets selection on the model regression prediction ability in BMGs is important but there is still no systematic and comprehensive study on it.

In this work, we focus on the effect of the data imbalance on the model prediction and attempt to provide solutions to avoid the results from trapping into the intensive training data area or from inducing by the data distribution of systems with similar composition. Ternary systems are chosen for study because they offer abundant data and avoid significant increasing complexity in chemical space compared with higher-order multicomponent systems. A machine learning regression algorithm that integrates multiple models to achieve predictions of the $D_{max}$ of MGs is established. And a serials of training sub-datasets containing different quantities of similar composition data for prediction are designed and their results are tested. By evaluating the prediction results of different sub-datasets, the strategy to prune the training dataset to model and predict various systems commencing from a similar information base is proved to mitigate the impact of data imbalance. To compensate the information loss caused by data reduction, we introduce an experimental iterative approach. Using this workflow, we validate its effectiveness through identifying the high GFA composition in the Zr-Al-Cu system by treating it as a completely unknown new system for prediction. Finally, the application of this workflow in the field of BMGs and the existing challenges are discussed.

## 2. Method

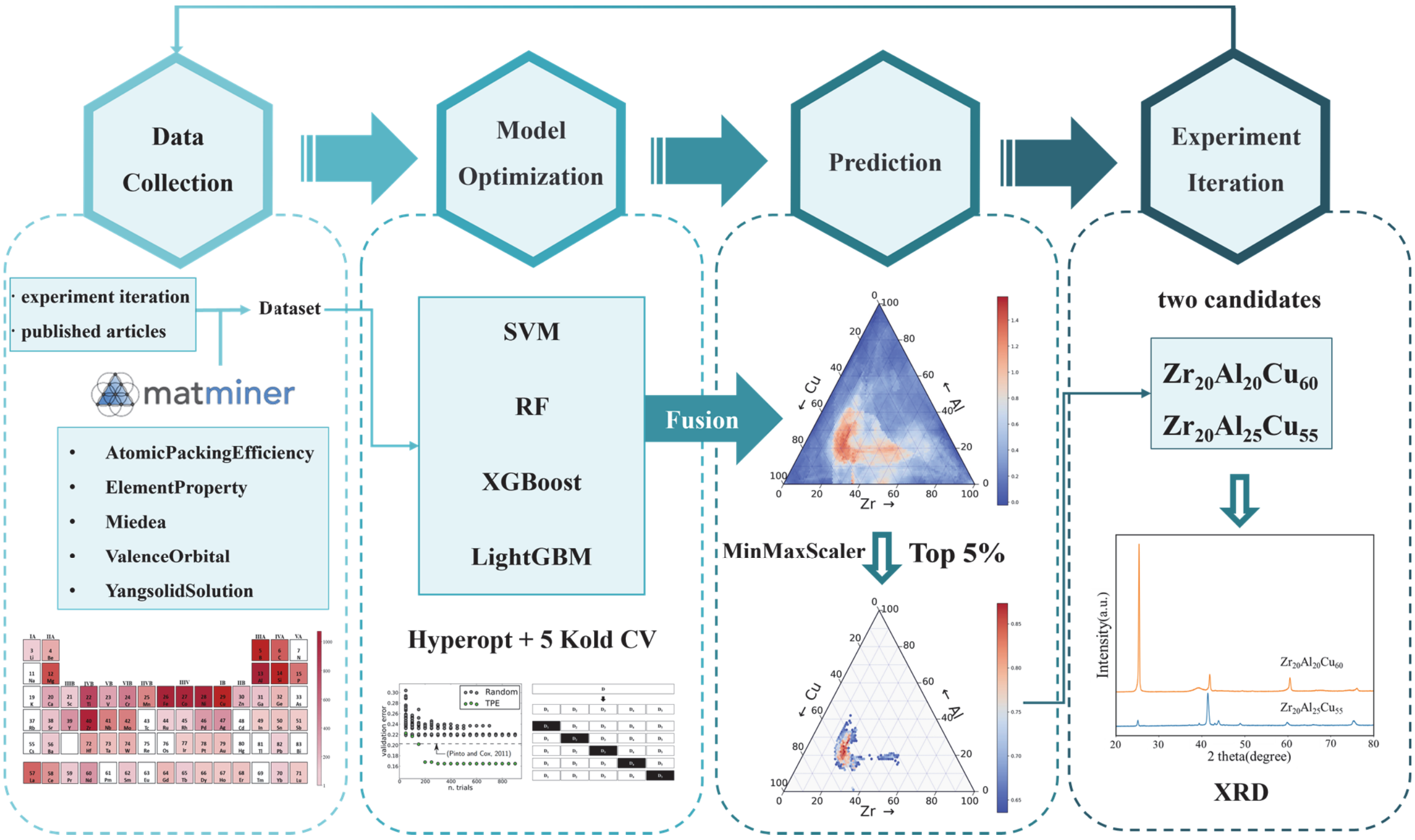

Fig.2 Machine learning and experimental iteration workflow proposed in this work. Models are trained and optimized using Hyperopt, with model fusion improving prediction performance. The top 5% points exhibiting high GFA are selected based on ranked $D_{max}$ prediction. Two candidates are experimentally validated in each iteration process and their results are added to the training set for the next iteration.

### *2.1 General workflow and model training*

As shown in Fig. 2, we export five groups of descriptors using the matminer Python library [38] to represent the physical and chemical properties of each composition. On this dataset, we initially train a set of models, including Support Vector Machine (SVM), Random Forest (RF), eXtreme Gradient Boosting (XGBoost), and Light Gradient Boosting (LightGBM). These models are classical machine learning models, which have good prediction ability for MGs and have certain anti-overfitting ability [39-41]. Based on the Hyperopt library, we choose the Tree-Structured Parzen Estimator (TPE) optimization algorithm and stop iterate when there is no improvement on the test set for 50 consecutive iterations in the 5-fold validation [42]. The selection range of hyperparameters is detailed in Table S1. Afterwards, we perform model fusion operation by using the four predicted values of sub-models as new inputs and optimizing the weights of these four values as hyperparameters. The final prediction result can be written as:

$$pred = \frac{x_1 \times y_{SVM} + x_2 \times y_{RF} + x_3 \times y_{XGBoost} + x_4 \times y_{LightGBM}}{x_1 + x_2 + x_3 + x_4} \quad (1)$$

where, x is the weight of each sub-model prediction result, which is optimized as a hyperparameter. y represents the prediction result output by each sub-model, and pred represents the result of the final output of the fusion model.

When the whole dataset is chosen for training, compared to the four individual models, the fusion model shows better prediction performance, with an $R^2$ value of 0.82 on the test set under the condition of 5-fold cross validation (as shown in Fig S1). For each ternary system, GFA predictions are conducted by varying the composition with a 1% atomic ratio change, resulting a total of 4851 composition predicted for each system. We normalize the predicted $D_{max}$ values to the range of [0,1], where a value closer to 1 indicates a higher $D_{max}$ prediction value in this system, while a value closer to 0 indicates a lower $D_{max}$ prediction value. The points corresponding to the top 5% highest predicted $D_{max}$ values of each system are selected as candidate composition and illustrated in the phase diagram. In the largest candidate area, we choose two candidates to synthesize in each experimental iteration for validation: one is the point with the highest predicted value, the other has the largest predicted value in a group of six points surrounding the first one. We prepare ribbon samples by melt-spinning, and prepare cylindrical BMG samples of different sizes by copper mold casting in molds of different diameters, starting from 1 mm diameter and increasing by 0.5 mm each time until the sample exhibits diffraction patterns containing crystalline structures in XRD testing. The obtained experimental results are subsequently incorporated as new training data into the dataset for the next iteration.

*2.2 Training data processing*

The $D_{max}$ of the composition that can be prepared as BMGs is taken as the original reported or experimental value, the $D_{max}$ of the composition that can be prepared as ribbon amorphous samples is taken as 0.2 mm [43], and the corresponding label value of the composition that cannot be obtained as amorphous samples is taken as 0 mm. Different reported values of the same composition are averaged and included in the database. Afterwards, the entire dataset will be pruned step by step by removing part of data that contributes to different forms of data imbalance, to test how to minimize the impact of data imbalance on the model's prediction ability.

*2.3 Descriptors selection*

For feature selection, we utilize the matminer Python library[38] to derive descriptors based on material composition. We incorporate descriptors from "ElementProperty" database and "ValenceOrbital" database to calculate the physical and chemical properties of each element, such as electronegativity, atomic radius, thermal conductivity, electron distribution, and so on. Additionally, we also consider other characteristics that may be related to GFA. Laws et al. [44] investigated the correlation between mean packing efficiency and GFA in their study, revealing that ideal atomic packing efficiency is likely to be one of the necessary conditions for high GFA. Therefore, based on their studies, we derive descriptors in this group to describe the atomic packing efficiency according to alloy compositions. We also derive solid solution phase stability descriptors based on Yang's work [45], in which multi-component BMGs are concentrated in the range of $\Omega$ (the entropy of mixing timing the average melting temperature of the elements over the enthalpy of mixing) less than 1.1 and $\gamma$ (the mean square deviation of the atomic size of elements) less than 5%, so we derive these two descriptors based on this article. Finally, we derive three descriptors based on the energy difference between crystalline and amorphous states using the semi-empirical model of Miedema et al [46]. A total of 86 descriptors are exported. Among them, 39 descriptors from the "ElementProperty" database are excluded due to their lack of variation with changes in elemental composition ratios within the same system. In addition, 6 descriptors with missing values are removed, and 12 descriptors are further removed due to strong linear correlations. Consequently, a refined dataset consisting of 29 descriptors (as shown in Table S2) is obtained through these filtration steps. The absolute value of the linear correlation between all pairwise descriptors is less than 0.8 (as shown in Fig S2), which means that there is no significant interdependence between this set of descriptors and can be used for the prediction of $D_{max}$.

## 3. Result & Discussion

### *3.1 Effect of imbalanced training datasets on prediction*

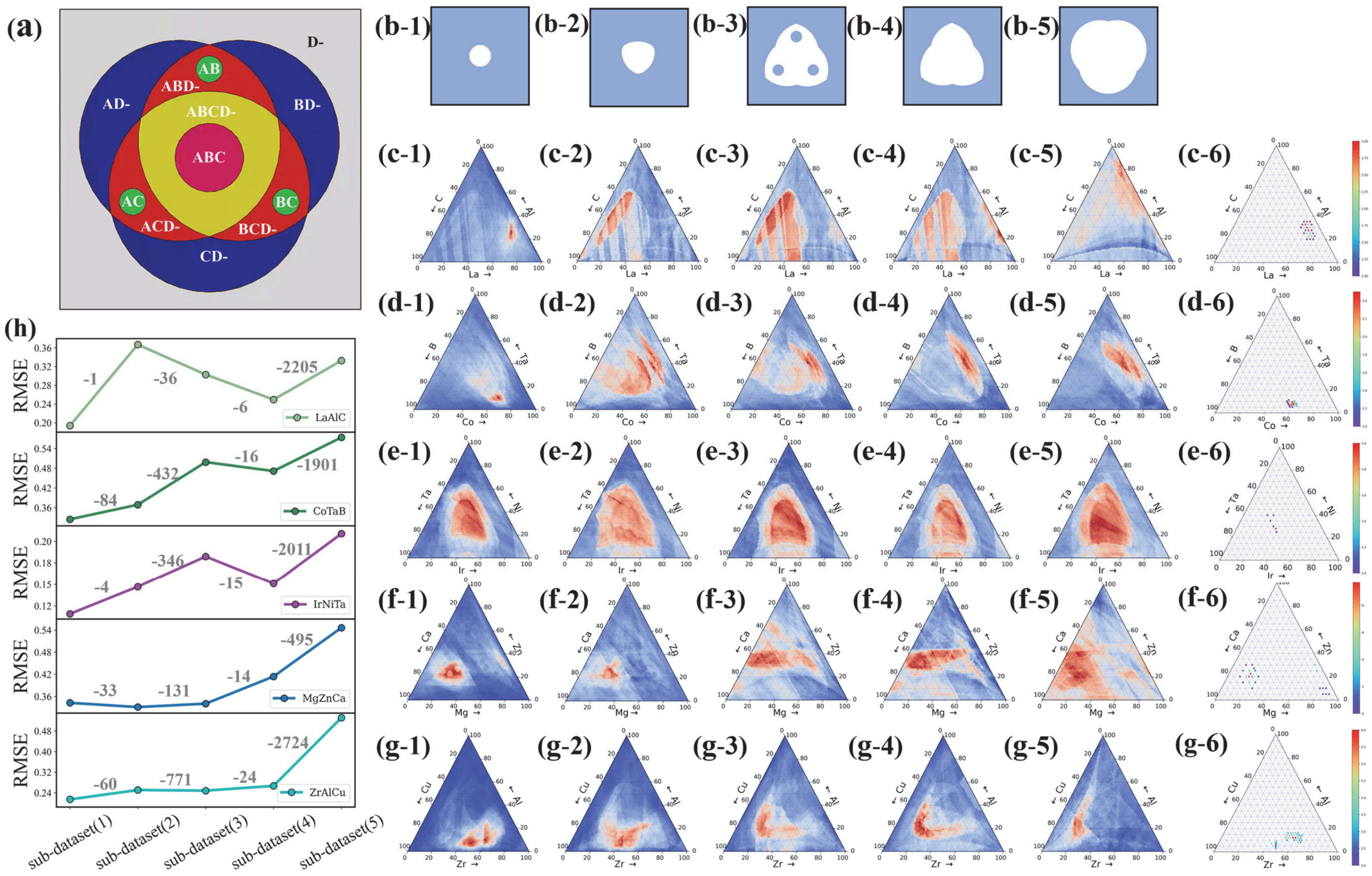

Fig. 3 The impact of varying degrees of data imbalance on the model prediction (a) diagram depicting the data type division in a ternary system; (b) diagrams illustrating the data range of the sub-datasets; (c~g) prediction results of La-Al-C, Co-Ta-B, Ir-Ni-Ta, Mg-Zn-Ca, Zr-Al-Cu, system ((x-1) ~ (x-5) represent the prediction results in the cases of (1) to (5) sub-datasets in turn and (x-6)

represents the true value of the experimental reports, where x= b, c, d, e, f) ;(h) effect of pruning relevant data on the error of model prediction results (the labelled numbers indicate the data entries reduced from the previous dataset).

Although as most of the reported machine learning models of BMGs, our optimized fusion model trained on the entire experimental dataset demonstrates prediction ability on the test set under the condition of 5-fold cross validation, its ability in systems which have not been experimentally investigated or in the composition areas which have sparse data is still questionable. Therefore, based on the aforementioned dataset, we design a series of sub-datasets to emulate the varying degrees of data loss and to explore the influence of data imbalance on the predictive ability of the model. Sub-datasets for the prediction target of A-B-C system (where A, B, C represent different elements) are designed and their prediction results are evaluated in Fig. 3. As shown in Fig 3(a), We classify the data containing varying degrees of information on the composition of A-B-C system into five categories as follows: (I) data for composition containing only one of the three elements, denoted as AD-, BD-, CD- ("D" represents elements other than A,B and C, and "D-" can represent more than one element); (II) binary data containing two of the three elements, namely AB, BC, AC; (III) data for alloys of three or more elements containing two of the three elements, namely ABD-, BCD-, ACD-; (IV) data for alloys of four or more elements containing these three elements, i.e., ABCD-; and (V) data representing the ternary alloy system itself, namely ABC. For example, in the case of the Zr-Al-Cu system as the target, $Ti_{45}Zr_{20}Be_{35}$, $Zr_{50}Cu_{50}$, $Zr_{40}Al_{20}Zn_{40}$, $Zr_{40}Al_{25}Cu_{25}Co_{10}$, $Zr_{60}Al_{10}Cu_{30}$ belong to the data category (I), (II), (III), (IV) and (V), respectively. From category (I) to (V), the data exhibit an increasing association with the target system. To explore their effect on the model prediction, five sub-datasets with varying degrees of missing relevant data are designed by reducing the data in category from (V) to (I) step by step, as shown in Fig. 3(b). In this way, the imbalance decreases from left to right for the datasets presented in Fig. 3(b). Five ternary systems, La-Al-C [34], Co-Ta-B [33,47,48], Ir-Ni-Ta [49],

Mg-Zn-Ca [50], Zr-Al-Cu [51-54] are trained based on above five categories sub-datasets. The model's prediction ability is compared among them and the results are shown in Fig. 3(c~g). To accurately assess the impact of data imbalance, we normalize the collected experimental data in each prediction to [0, 1]. The prediction ability is quantified through the relative mean squared error (RMSE), as presented in Fig. 3 (h).

It is evident that the model's predicted results in sub-dataset(1) show a significant overlap with the high GFA region reported in collected experimental data. In Fig. 3(h), predicted results in sub-datasets(1) generally have the smallest RMSE. This is mainly because the ABCD-type data present in sub-datasets(1) are generally developed from the ABC amorphous composition thus they have a strong similarity to the target system. In the La-Al-C system, a significant increase of the RMSE is observed with the deletion of only one data point, as shown in Fig. 3(c-1) and 3(c-2). The composition corresponding to this data point is $La_{60}Al_{26}Ga_{4}C_{10}$, where the Ga element is adjacent to Al in the periodic table and this data point is therefore close to $La_{60}Al_{30}C_{10}$ in feature space, playing a strong guiding impact to trap the predicted high $D_{max}$ value near to this composition. In the subsequent data gradient change, the regions of high GFA composition in the predicted results of La-Al-C and Co-Ta-B systems are almost unchanged, as shown in Fig. 3(c) and 3(d). For Ir-Ni-Ta system, as shown in Fig. 3(e), the variation of high GFA regions is insignificant in several cases. Nevertheless, for the Mg-Zn-Ca and Zr-Al-Cu systems the impact of data imbalance become notable starting from sub-dataset(4) where the RMSE increases obviously as shown in Fig 3(h). The inclusion of category (III) data in the training set causes the predicted high GFA regions aggregating to the experiment reported high GFA regions, resulting the prediction results being biased towards the data-intensive regions in the training dataset. The bias caused by the data imbalance poses challenges for the discovery of new BMGs composition. For example, in the Mg-Zn-Ca system, there are two high GFA composition with significant differences in composition. To be specific, both $Mg_{20}Zn_{20}Ca_{60}$ in the low Mg region and $Mg_{70}Ca_{5}Zn_{25}$ in the high Mg region have the same $D_{max}$ value. However, in Fig. 3(g-2), only one high $D_{max}$ region can be observed, and it is only after removing category (III) data that a second high $D_{max}$ prediction region can be observed in Fig. 3(f-3). This imbalance in the data is likely to result in the overlooking of potential high GFA regions, thus significantly compromising the effectiveness of machine learning on looking for new composition with candidate properties.

To mitigate the undesirable influence of a large amount of relevant experimental data on the prediction results, we choose the training set selection strategy of sub-dataset(3) for further discussion. In this final selected sub-dataset, there is no data that is highly similar to the element composition of the ternary system to be predicted. Taking the Zr-Al-Cu system as an example, the sub-dataset used at this time does not contain the ternary data containing two of the three elements of Zr, Al and Cu, which reflects the unbiasedness. This selection criteria serves a dual purpose. On one hand, it helps reduce the adverse guiding effect arising from similar composition data. On the other hand, it enables us to model composition exploration in unfamiliar systems, where data category (III) to (IV) for the target system are unavailable. Thus, each system is predicted as an unexplored new system. To compensate for the information loss resulting from data reduction, we next conduct a small number of experiments based on the prediction results and iterate the experimental data into the machine learning model for another round of predictions, aiming to identify the high GFA area in the target system even it is regarded as an entirely unexplored new system [28,55].

*3.2 Experimental iteration*

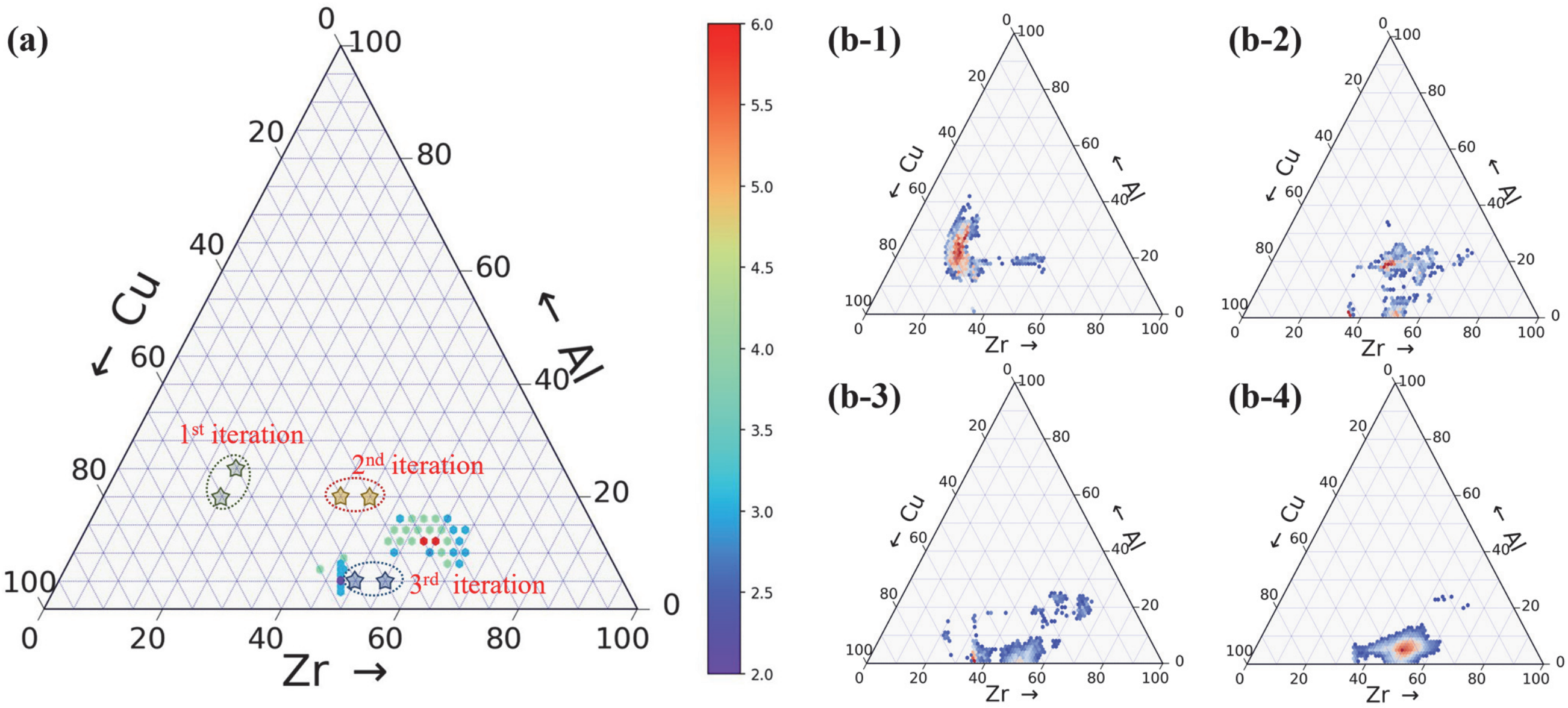

Fig. 4 Iterative prediction empowered by adding experimental data step by step. (a) reported $R_c$ for Zr-Al-Cu system, where pentagrams correspond to the iteration data points, (b)iterative prediction results for Zr-Al-Cu system.

We choose Zr-Al-Cu system to validate the iteration scheme between experiments and machine learning models, due to its relatively simple preparation process, low cost, as well as abundant experimental data available for comparison. As shown in Fig. 4(a), the circled pentagrams correspond to the data points that are experimentally validated and used for each iteration. Fig. 4(b-1) shows the top 5% region of the $D_{max}$ in the initial model prediction without any iteration, where the color represents the relative predicted value. Fig. 4(b-2~b-4) displays the prediction results after the first, second, and third iterations, respectively. According to the prediction results in Fig. 4(b-1), two candidates, $Zr_{20}Al_{20}Cu_{60}$ and $Zr_{20}Al_{25}Cu_{55}$ with a difference of five atomic percentage in composition are chosen for experimental verification and iteration. We explore the $D_{max}$ of candidates by copper casting and melt spin quenching. The experimental results (see in Fig. S3) indicate that $Zr_{20}Al_{20}Cu_{60}$ and $Zr_{20}Al_{25}Cu_{55}$ cannot be prepared as amorphous samples, and the XRD patterns of the ribbon samples obtained through melt spin quenching exhibit obvious crystalline structures. The two experimental data are added into the training set and a new model is obtained. Using the new model, the high

GFA area moves to regions with higher Zr concentration as shown in Fig. 4(b-2). In the second iteration, $Zr_{40}Al_{20}Cu_{40}$ and $Zr_{45}Al_{20}Cu_{35}$ are chosen for experimental validation, and ribbon amorphous samples are successfully yielded for them. Furthermore, Fig. 4(b-3) is the results predicted by the next trained model with two more experimental data added. The third iteration data, $Zr_{50}Al_5Cu_{45}$ and $Zr_{55}Al_5Cu_{40}$, can obtain cylindrical amorphous samples with diameters of 2.5 mm and 1.5 mm, respectively. As depicted in Fig. 4(b-4), the highest point of the predicted result coincides with the experimental iteration point, and is close to the experimental reported highest $D_{max}$ at composition within the distance of 10 atomic percentage. It can be observed that as we add experimental data, the prediction results gradually approach the amorphous area. During the experimental process, the GFA corresponding to the selected iteration data becomes higher and higher, confirming that iteration can indeed improve the prediction effectiveness of the model.

Additionally, we examine the high GFA region centered around $Zr_{50}Al_5Cu_{45}$ in Fig. 4(b-4). We proceed to select boundary points with equal Zr content, equal Al content, and equal Cu content, respectively, using increments of five atomic ratios to conduct experiments in order to verify the potential formation of BMGs in this area. As shown in Fig. S3, the 1 mm diameter rod-like samples of $Zr_{45}Al_{10}Cu_{45}$ and $Zr_{35}Al_5Cu_{60}$ proved to be amorphous, the other two candidates can be prepared into ribbon amorphous samples. This indicates that, the pruned dataset effectively mitigates the data imbalance issue and iteration scheme with small amount of experiment exploration help to identify a relatively high GFA region in Zr-Al-Cu systems even there is no data of the ternary system provided beforehand. Training and predicting each system as an unexplored one by reducing the strong related data from the dataset decrease the chance to be trapped into the known data-intensive area and will benefit to find new composition with high GFA.

## 4. Conclusion

In this work, we analyze the origin and effect of data imbalance in the datasets of BMGs. The way to mitigate the effect of this imbalance is proposed by testing the model prediction ability using varying pruned degrees of the datasets. And the sub-dataset without ABC-related information is chosen since it avoids the results trapped at the data-intensive area. By combining the machine learning and experimental iteration, we mitigate the negative impact of data imbalance and accelerate the search for high GFA regions, especially suitable for discovery new composition in unexplored system and beyond experimental intensive data area in known systems.

The main conclusions are as follows:

(1) By designing multiple sub-datasets with gradients of similarity to the elements of the target system, we demonstrate that imbalanced data distribution can introduce bias, which has a strong guiding impact on the predictive results of the model. An imbalanced data set can result in biased predictions in densely populated areas, reducing the applicability of the model. By restricting the training sets exclude the ABC-related experimental data, all the target system can be predicted at almost the same level, no matter how many parts of the specific chemical space have not been studied yet. It is important to note that this does not negate the effectiveness of adding data in an even manner, as doing so can improve the prediction accuracy of the model to some extent. It is suitable for integration with high-throughput BMGs preparation methods such as laser melting [56-58] to speed up the high-quality data production in BMGs exploration.

(2) We propose a workflow that involves a small number of experiments and iterative model training to identify high GFA regions in a specified system. This combination of machine learning and experiment methods can help researchers to identify regions with relatively higher GFA in the Zr-Al-Cu system. Additionally, this method can be expanded to predict other properties of amorphous alloys, effectively mitigating the impact of data imbalance and reducing the time required for experimentation. By simply substituting the labels in the dataset, we can extend the predictive target to other properties of interest in the field of MGs, such as glass transition temperature or elastic modulus. The success of above strategy lies in two aspects. Firstly, by excluding the ABC-related data from the training dataset, the prediction values are prevented from being biased towards the dense region of training data which mostly only contains amorphous cases. Secondly, the data iteratively added through experiments includes both crystalline and amorphous results, continuously improving the prediction across the whole chemical space.

Theoretically, the computational and experimental costs increase when targeting a quaternary or higher system. The composition space to be explored becomes more intricate, leading to a broader range of iteration choices. Nevertheless, when compared with the conventional “trial-and-error” approach for exploring new compositions, the method outlined in this paper demonstrates notable efficiency advantages within the constraints of comparable complexity. This approach will expedite the development of new BMGs compositions, especially for unexplored systems.

**Appendix A. Data and code availability**

All the data and codes for fusion model can be accessed via https://github.com/xuhegg/ML_BMG_GFA.

**Appendix B. Supplementary data**

Supplementary data to this article can be found online.

**Acknowledgements**

This study was funded by the National Natural Science Foundation of China (Grant No. 52171150), the Informatization Plan of Chinese Academy of Sciences (Grant No. CAS-WX2021SF-0102), and the National Key Research and Development Program of China (Grant No. 2018YFA0703601). The funder played no role in study design, data collection, analysis and interpretation of data, or the writing of this manuscript.

**Author contributions**

R.L. designed this work. R.L. and R.X. guided the completion of the method. X.G. and R.X. constructed the models, X.G. conducted the materials synthesis and model iterations. J.B. and X.L. collected and organized the dataset. All the authors participated in the analysis of the data and discussions of the results, as well as in preparing the paper.

## Competing interests

All authors declare no financial or non-financial competing interests.

# Efficient Learning Strategy for Predicting Glass Forming Ability in Imbalanced Datasets of Bulk Metallic Glasses

Xuhe Gong [a,b], Jiazi Bi [a], Xiaobin Liu [a], Ran Li[a,*], Ruijuan Xiao [b,*], Tao Zhang [a], Hong Li [b]

[a] *School of Materials Science and Engineering, Key Laboratory of Aerospace Materials and Performance (Ministry of Education), Beihang University, Beijing 100191, China;*

[b] *Institute of Physics, Chinese Academy of Sciences, Beijing 100190, China*

*E-mail: liran@buaa.edu.cn (R. Li), rjxiao@aphy.iphy.ac.cn (R. Xiao).

## Supplementary Information

**Table of Contents**

| Table S1. Hyperparameter range | |
|---|---|
| Model | Hyperparameters range (start, end, step) |
| SVM | “C”: (1,50,1)<br>“gamma”: (0.1,0.45,0.005)<br>“epsilon”: (0,0.2,0.002) |
| RF | “n_estimators”: (50,550,5)<br>“max_features”: (3,29,1)<br>“max_depth”: (8,55,1)<br>“min_samples_split”: (2,10,1)<br>“min_impurity_decrease”: (0,5,0.1) |
| XGBoost | "n_estimators": (150,450,3)<br>"learning_rate": (0.05,0.3,0.002)<br>"colsample_bytree": (0.3,1,0.1)<br>"colsample_bynode": (0.1,1,0.1)<br>"gamma": (0,15,0.2)<br>"reg_lambda": (0,25,0.5)<br>"min_child_weight": (0,50,0.5)<br>"max_depth": (5,45,1) |
| LightGBM | “n_estimators”: (100,800,5)<br>“num_leaves”: (10,400,5)<br>“learning_rate”: (0.1,0.5,0.02)<br>“min_child_samples”: (1,40,1)<br>“reg_alpha”: (0,10,0.5)<br>"reg_lambda”: (0,100,2) |

| Table S2. Descriptors in the dataset and their interpretation | |
|---|---|
| descriptors | explanation |
| mean simul. packing efficiency | The pacsking efficiency is based on the Atomic Packing Efficiency (APE), which measures the difference between the ratio of the radii of the central atom to its neighbors and the ideal ratio of a cluster with the same number of atoms that has optimal packing efficiency. |
| mean abs simul. packing efficiency | |
| dist from 1 cluster $\lvert APE \rvert < 0.010$ | the distance between an alloy composition and the 1 clusters that have a packing efficiency below 0.010 from ideal |
| mean X | mean value of electronegativity from Pymatgen data |
| std_dev X | standard deviation of electronegativity from Pymatgen data |
| mean Group | mean value of group from Pymatgen data |
| std_dev Group | standard deviation of group from Pymatgen data |
| mean Block | mean value of block from Pymatgen data |
| std_dev Block | standard deviation of block from Pymatgen data |
| mean atomic_mass | mean value of atomic mass from Pymatgen data |
| std_dev atomic_mass | standard deviation of atomic mass from Pymatgen data |
| mean atomic_radius | mean value of atomic radius from Pymatgen data |
| std_dev atomic_radius | standard deviation of atomic radius from Pymatgen data |
| mean mendeleev_no | mean value of Mendeleev number from Pymatgen data |
| std_dev mendeleev_no | standard deviation of Mendeleev number from Pymatgen data |
| mean electrical_resistivity | mean value of electrical resistivity from Pymatgen data |
| mean thermal_conductivity | mean value of thermal conductivity from Pymatgen data |
| std_dev thermal_conductivity | standard deviation of thermal conductivity from Pymatgen data |
| mean melting_point | mean value of melting point from Pymatgen data |
| std_dev melting_point | standard deviation of melting point from Pymatgen data |
| Miedema_deltaH_amor | the formation enthalpies of the intermetallic compound, solid solution and amorphous phase of a given composition, based on semi-empirical Miedema model |
| Miedema_deltaH_ami | |
| Miedema_deltaH_am_min | Miedema_deltaH_ami = Miedema_deltaH_amor - Miedema_deltaH_inter<br>Miedema_deltaH_ams = Miedema_deltaH_amor - Miedema_deltaH_ss_min<br>Miedema_deltaH_am_min = min(Miedema_deltaH_ami, Miedema_deltaH_ams) |
| frac s valence electrons | fraction of valence electrons in s orbitals |
| frac p valence electrons | fraction of valence electrons in p orbitals |
| frac d valence electrons | fraction of valence electrons in d orbitals |
| frac f valence electrons | fraction of valence electrons in f orbitals |
| Yang omega | $$\Omega = \frac{T_m \Delta S_{mix}}{\lvert \Delta H_{mix} \rvert}$$ where, $T_m$ represents the melting point, $\Delta S_{mix}$ represents the entropy change of mixing, and $\Delta H_{mix}$ represents the enthalpy change of mixing |
| Yang delta | $$\delta = \sqrt{\sum_{i=1}^{n} c_i (1 - \frac{r_i}{\bar{r}})^2}$$ where, $c_i$ represents the element proportion, $r_i$ represents the atomic radius of the i-th element, and $\bar{r}$ represents the average radius of the element in alloy |

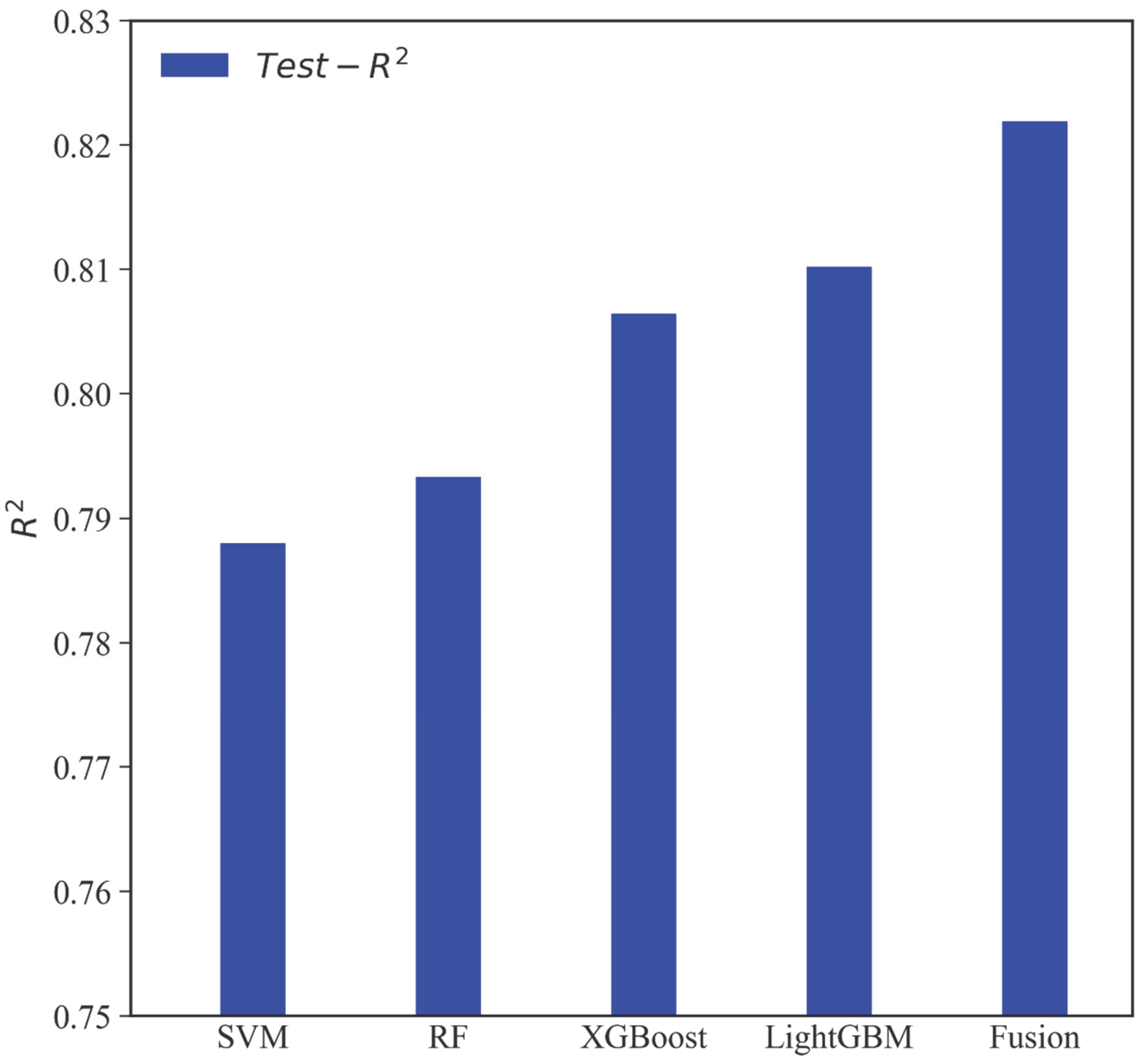

Fig. S1 Performances of the machine learning algorithms. In this article, we measure the prediction accuracy of the model by $R^2$ on the five-fold cross-validation test set. The prediction effects of five models, SVM, RF, XGBoost, LightGBM, and fusion model are shown in turn in the figure, and the fusion model performs better in comparison.

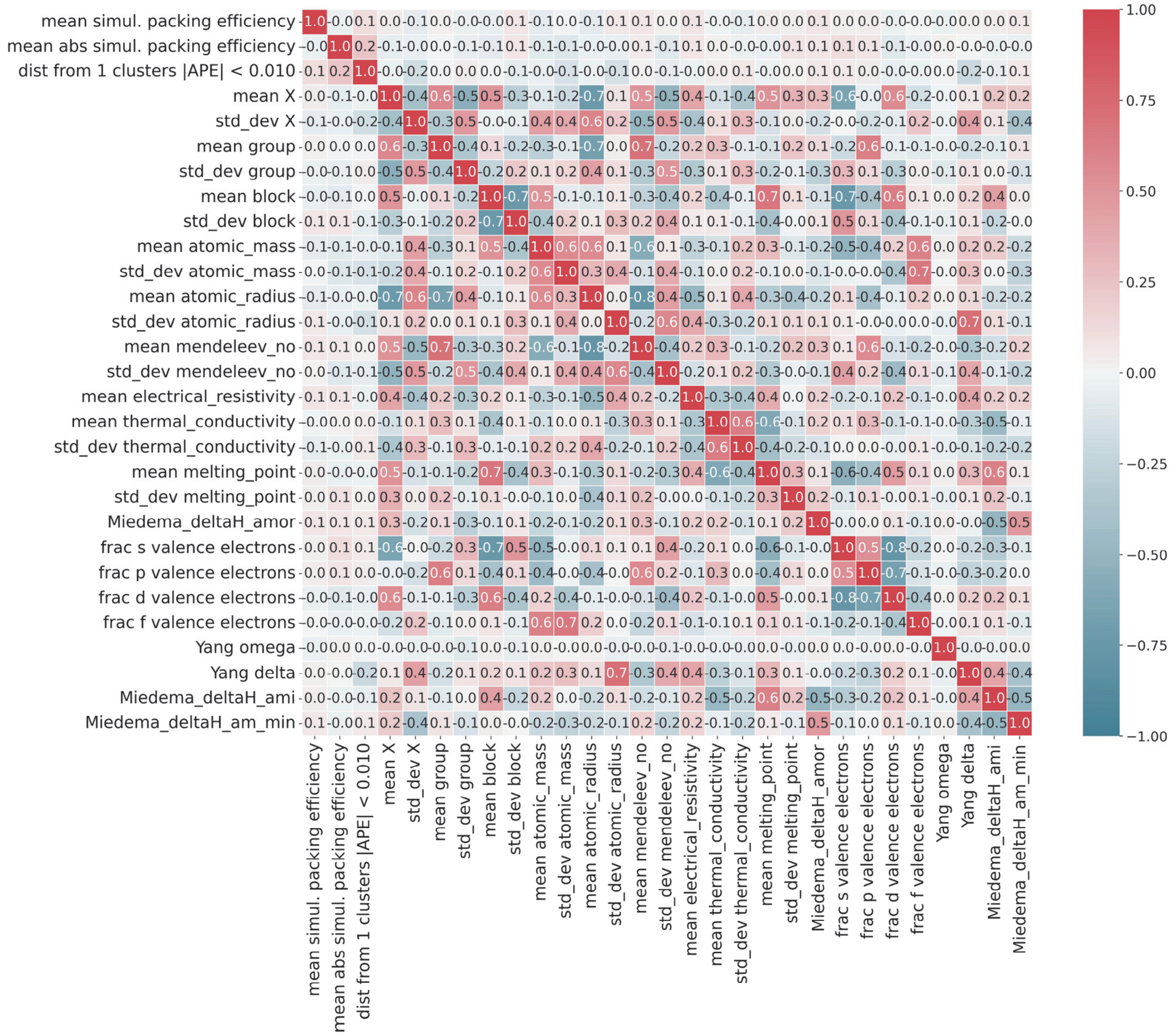

Fig. S2 Feature linear correlation matrix.

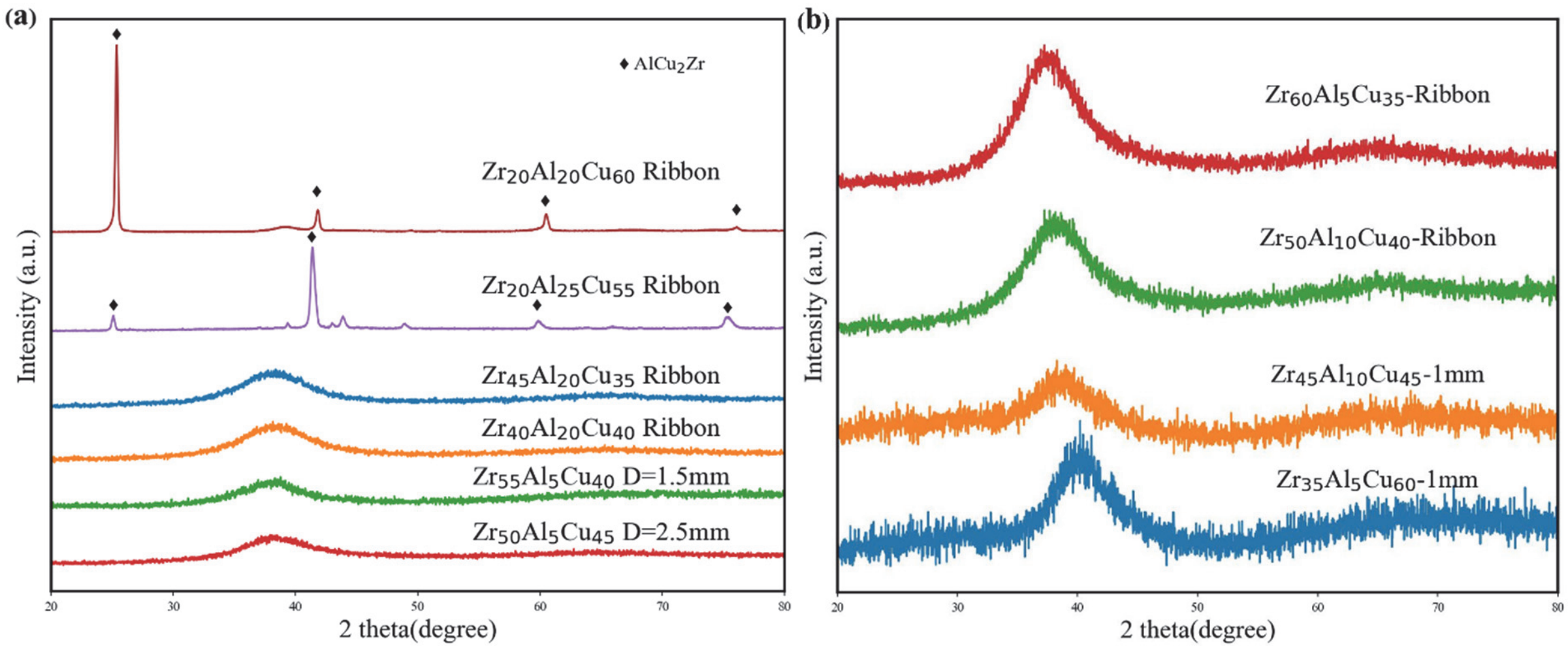

Fig. S3 XRD patterns of the experimental iterative data. (a) The experimental iteration data of BMGs composition search in Zr-Al-Cu system, from top to bottom, are the diffraction curves of the first, second and third iteration data. (b) After three rounds of iteration, the boundary GFA verification of the range selected by the top 5% predicted value shows that two of the four boundary points can form bulk amorphous samples and two can form ribbon amorphous samples.